# Fast solution of the superconducting dynamo benchmark problem

**Leonid Prigozhin[1] and Vladimir Sokolovsky[2]**

[1]J. Blaustein Institutes for Desert Research, Ben-Gurion University of the Negev, Sde Boqer Campus 84990, Israel
[2]Physics Department, Ben-Gurion University of the Negev, Beer-Sheva 84105, Israel

E-mail: leonid@bgu.ac.il and sokolovv@bgu.ac.il,

**Abstract**
A model of high temperature superconducting dynamo, a promising type of flux pumps capable of wireless injection of a large DC current into a superconducting circuit, has recently been chosen as an applied superconductivity benchmark problem and solved using ten different numerical methods (Ainslie *et al* 2020 *Supercond. Sci. Technol.* **33** 105009). Using expansions in Chebyshev polynomials for approximation in space and the method of lines for integration in time we derive a simple and accurate numerical method which is much faster. The proposed numerical method was applied also to problems with transport current and a field-dependent sheet critical current density.

## 1. Introduction

The high-$T_c$ superconducting (HTS) dynamos are devices capable of inducing a large DC current in a superconducting circuit without the cryogenic losses associated with non-superconducting current leads [1-5]. Mathematical models, explaining the voltage in an HTS strip (the dynamo stator) generated by the dynamo-rotor-attached permanent magnet (PM), have been proposed recently, see [6-10] and the references therein. Available mathematical descriptions are either simplified circuit-type models of the HTS dynamo (e.g. [6]) or consider an infinite superconducting strip subjected to the time-varying non-uniform field of a rotating long PM [7-10]. Models of the latter type are less simplified and better reproduce the non-trivial features of the HTS dynamos.

Such HTS dynamo model, with a coated conductor stator characterized by the nonlinear current-voltage relation with a field-independent critical current density, has been proposed as a new benchmark problem in applied superconductivity and solved by several numerical methods in [11]. These authors computed the generated open-circuit voltage, compared the computation times needed to the methods employed, and made their simulation results available as a supplementary material.

Here we show that an accurate numerical solution of this benchmark problem can be obtained using Chebyshev polynomials for the approximation in space. Resulting semi-discretized problem, a system of ordinary differential equations (ODE), can then be integrated in time using a standard ODE solver with an automatic choice of the time step. Such discretization scheme (the method of lines) helps to control the accuracy of integration. For HTS strips, strip stacks, and pancake coils a similar approach has been proposed in [12]. As in [12], our choice of Chebyshev polynomials for spacial approximation is based on two factors: remarkably fast convergence of the interpolation by Chebyshev polynomial expansions [13] and the ease of using such expansions for solving integral equations with a singular Cauchy-type kernel (see, e.g., [14, 15] ). The derived method is much faster than all methods in [11].

Numerical simulations in our work were done using Matlab R2020a on a PC with the Intel (R) Core (TM) i7-9700 CPU@ 3,00 Hz and 32 GB RAM.

## 2. The benchmark problem

Following [11], we assume that an infinitely long PM with the remanent flux density $B_r$ rotates counter clockwise past a thin stationary HTS strip in the open-circuit configuration (Figure 1); all HTS dynamo parameters are listed in Table 1.

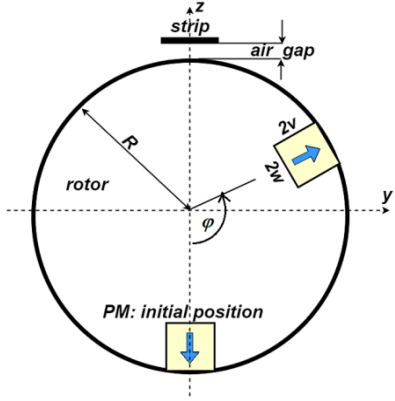

Figure 1. A scheme of an HTS dynamo: the geometry of the benchmark problem.

Table 1. HTS dynamo benchmark parameters (as in [11]).

| | | |
|---|---|---|
| permanent magnet (PM) | width, $2w$ | 6 mm |
| | height, $2v$ | 12 mm |
| | active length, $l$ | 12.7 mm |
| | remanent flux density, $B_r$ | 1.25 T |
| HTS stator strip | width, $2a$ | 12 mm |
| | thickness | 1 μm |
| | critical current, $I_c$ | 283 A |
| | power value, $n$ | 20 |
| rotor external radius, $R$ | | 35 mm |
| air gap, $d$ | | 3.7 mm |
| frequency of rotation, $f$ | | 4.25 Hz |

If an infinitely long permanent magnet is in the initial position (Figure 1), its magnetic field $\boldsymbol{H}_0 = (H_{0,y}, H_{0,z})^T$ at a point $\boldsymbol{p} = (y, z)^T$ outside the magnet is, see [16],

$$H_{0,y}(\boldsymbol{p}) = -\frac{B_r}{4\pi\mu_0}\left\{\ln\left[\frac{(y+w)^2+(\tilde{z}-v)^2}{(y+w)^2+(\tilde{z}+v)^2}\right] - \ln\left[\frac{(y-w)^2+(\tilde{z}-v)^2}{(y-w)^2+(\tilde{z}+v)^2}\right]\right\},$$

$$H_{0,z}(\boldsymbol{p}) = -\frac{B_r}{2\pi\mu_0}\left\{\arctan\left[\frac{2v(y+w)}{(y+w)^2+\tilde{z}^2-v^2}\right] - \arctan\left[\frac{2v(y-w)}{(y-w)^2+\tilde{z}^2-v^2}\right]\right\},$$

where $\mu_0$ is the permeability of vacuum, $\tilde{z} = z + R - v$, $R$ is the rotor radius, $2v$ and $2w$ are the PM height and width, respectively. After rotation of the rotor by an angle $\varphi$ the field at the point $\boldsymbol{p}$ becomes

$$\boldsymbol{H}_{PM}(\varphi, \boldsymbol{p}) = A(\varphi)\boldsymbol{H}_0(A(-\varphi)\boldsymbol{p}), \tag{1}$$

where

$$A(\varphi) = \begin{pmatrix} \cos\varphi & \sin\varphi \\ -\sin\varphi & \cos\varphi \end{pmatrix}$$

is the matrix of rotation.

Neglecting the strip thickness, we present the strip as $\{-\infty < x < \infty, -a \leq y \leq a, z = z_s\}$, where $2a$ is the strip width, $z_s = R + d$, and $d$ is the air gap. It is assumed the parallel-to-strip electric field component $E$ and the strip sheet current density $J$ are directed along the strip (parallel to the $x$-axis) and obey the power law,

$$E = E_0 \left|\frac{J}{J_c}\right|^{n-1} \frac{J}{J_c}. \tag{2}$$

Here, as in [11], the field-independent sheet critical current density is $J_c = I_c/2a$, where $I_c$ is the strip critical current, $E_0 = 10^{-4}$ V/m, the power $n$ is a constant. By the Faraday law, at the strip points $\boldsymbol{p}_s = (y, z_s)^T$ we have

$$\frac{\partial E}{\partial y} = \mu_0 \frac{\partial H_z}{\partial t}. \tag{3}$$

Here the normal-to-strip component $H_z$ of the total magnetic field is a superposition of the PM and the strip current field components:

$$H_z(t, y) = H_{\text{PM},z}(\varphi(t), \boldsymbol{p}_s) + \frac{1}{2\pi} \int_{-a}^{a} \frac{J(t, u)}{y - u} du. \tag{4}$$

Since for all strip points $z = z_s$, to simplify our notations we will omit the dependence on $z$ and write $H_{\text{PM},z}(\varphi(t), y)$ instead of $H_{\text{PM},z}(\varphi(t), \boldsymbol{p}_s)$. Finally, if a transport current is given, the additional condition is

$$\int_{-a}^{a} J(t, y) dy = I(t), \tag{5}$$

where $I(t)$ is a known function. The open-circuit condition in the benchmark problem means the transport current $I$ is zero. Jointly with the initial condition $J(0, y) = 0$, equations (1)-(5) fully describe the HTS dynamo benchmark problem [11].

Experimentally, the instantaneous open-circuit voltage at the ends of a stator strip segment has been measured in [2, 7, 9, 10]. Taken for a segment containing the close-to-rotor active zone of the stator, this voltage and its time averaged value are the most important characteristics of an HTS dynamo. In the benchmark model the voltage is suggested to calculate as

$$V(t) = -\frac{l}{2a} \int_{-a}^{a} E(t, y) dy, \tag{6}$$

where $l$ is the length of the active zone (Table 1).

### 3. Numerical scheme

Rescaling the variables,

$$J' = J/J_c, \quad (H'_y, H'_z) = (H_y, H_z)/J_c, \quad E' = E/E_0,$$

$$(y', z') = (y, z)/a, \quad t' = \frac{E_0}{\mu_0 J_c a} t, \quad I' = \frac{I}{2aJ_c},$$

omitting the primes, and differentiating equations (4) and (5) with respect to time, we obtain in dimensionless form for $t > 0$, $-1 < y < 1$:

$$E = |J|^{n-1} J, \tag{7}$$

$$\frac{1}{2\pi} \int_{-1}^{1} \frac{\dot{J}(t, u)}{y - u} du = \frac{\partial E}{\partial y} - \dot{H}_{\text{PM},z}, \tag{8}$$

$$\int_{-1}^{1} \dot{J}(t,y)\,dy = 2\dot{I}(t). \tag{9}$$

Here the dot above variables means time derivative, the singular integral is understood in the principal value sense.

As in [12], we seek an approximation to the sheet current density in the form of a weighted expansion

$$J(t,y) = \frac{\sum_{i=0}^{N} \alpha_i(t) T_i(y)}{\sqrt{1-y^2}}, \tag{10}$$

where $T_i$ are Chebyshev polynomials of the first kind and $\alpha_i$ the expansion coefficients. Then we derive a system of ODE for $N$ unknowns, the values $J_k(t) = J(t, y_k)$ in $N$ mesh points, the roots of the Chebyshev polynomial $T_N(y) = \cos[N \arccos(y)]$:

$$y_k = \cos\left(\frac{\pi}{N}\left[k - \frac{1}{2}\right]\right), \quad k = 1, 2, \ldots, N. \tag{11}$$

Such meshes are denser near the interval ends and suppress the Runge phenomenon, the instability of polynomial interpolation on uniform meshes [13]. Substituting (10) into equation (8) we use the well-known relation [17]

$$\int_{-1}^{1} \frac{T_m(u)\,du}{(y-u)\sqrt{1-u^2}} = -\pi U_{m-1}(y), \tag{12}$$

where $y \in [-1,1]$, $U_{m-1}(y) = \sin(m \arccos y)/\sin(\arccos y)$ for $m = 1, 2, \ldots$ are Chebyshev polynomials of the second kind, and $U_{-1} = 0$. This yields, for $k = 1, 2, \ldots, N$,

$$-\frac{1}{2}\sum_{i=1}^{N} \dot{\alpha}_i U_{i-1}(y_k) = \left(\frac{\partial E}{\partial y} - \dot{H}_{PM,z}\right)\bigg|_{y_k}. \tag{13}$$

Substituting (10) also into equation (9) and noting that

$$\int_{-1}^{1} \frac{T_m(y)\,dy}{\sqrt{1-y^2}} = \begin{cases} 0 & m \neq 0, \\ \pi & m = 0, \end{cases}$$

we find $\dot{\alpha}_0 = 2\dot{I}(t)/\pi$. This simple equation ensures the total current constraint (5). Satisfying this constraint using other numerical methods is typically more complicated and needs iterations or adding a penalty term.

Let the $J_k$ values be known at time $t$. To find the right hand side of equations (13) we calculate the electric field (7) at the mesh nodes, $E_k = |J_k|^{n-1} J_k$, and, approximately, find $\partial E/\partial y$ at the same points by differentiating the Chebyshev expansion interpolating the $E_k$ values (see Appendix A). Furthermore,

$$\dot{H}_{PM,z} = \frac{\partial H_{PM,z}}{\partial \varphi}\frac{d\varphi}{dt} = 2\pi f \frac{\partial H_{PM,z}}{\partial \varphi}.$$

Computing $\partial H_{PM,z}/\partial \varphi$ for $\varphi = \varphi(t)$ at the mesh points $y_k$ (Appendix B), we calculate

$$\psi_k(t) = \left(\partial E/\partial y - \dot{H}_{PM,z}\right)\bigg|_{y=y_k}. \tag{14}$$

It is not difficult to construct the expansion $\psi$ in Chebyshev polynomials of the second kind,

$$\psi = \sum_{m=0}^{N-1} \beta_m U_m(y),$$

such that $\psi(y_k) = \psi_k$ (Appendix A). Equations (13) yield then $\dot{\alpha}_i = -2\beta_{i-1}$, $i = 1, ..., N$.

Finally, we set

$$\dot{J}_k(t) = \frac{\sum_{i=0}^{N} \dot{\alpha}_i(t) T_i(y_k)}{\sqrt{1 - y_k^2}}. \tag{15}$$

The ODE system (15) should be integrated in time. In our work we employed a standard Matlab ODE solver, *ode15s*, with the default parameters.

## 4. Solution of the benchmark problem

To compare our numerical simulation results to those in [11] we return to dimensional variables. As in [11], we assumed zero transport current, simulated ten full rotor rotations, and presented the $V(t)$ curves for the second cycle to avoid the transient dynamics during the first cycle.

First, to estimate the accuracy and convergence rate of our scheme, we solved the problem on different meshes and found that, visually, solutions for $N = 100$ and $N = 1000$ coincide (Figure 2). The sheet current density evolution during the second cycle is presented in Figure 3.

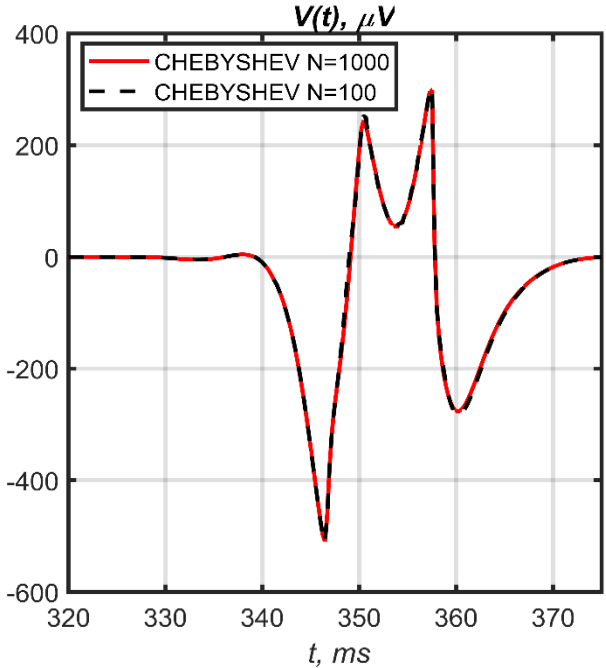

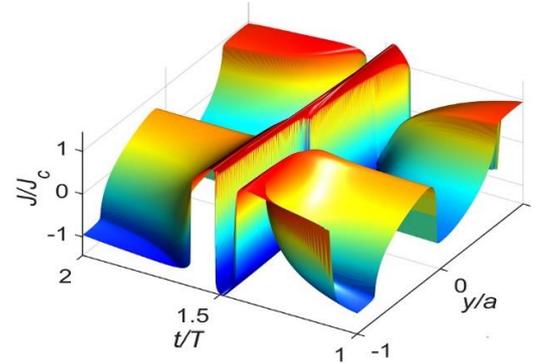

Figure 3. Sheet current density evolution in the second cycle, $T = 1/f$.

Figure 2. Dynamo generated voltage. Numerical solutions for 100 and 1000 mesh nodes.

Using the supplementary data to [11,18], we found that the $V(t)$ curves obtained using other methods are close. Here (Figure 4) we compare our solution to that of the fastest method in [11], called there MEMEP (the computations by this method have been corrected in [18]). Small differences, observed in this figure near the voltage peaks, are, probably, caused by different approaches to external magnetic field calculation (analytical in our work and numerical in [11]). Ignoring these differences, we can compare convergence of the MEMEP and our method (Table 2).

For this we calculate the $L^1$ norm, $\delta^N$, of the difference between the voltage $V(t)$ computed with an $N$-node mesh and that obtained using the same method with the finest mesh (one thousand nodes). For the MEMEP method the necessary data were obtained from [19]. The computation times, presented for similar computers in Table 2, show that similar accuracy can be achieved faster using our method. For other, much slower methods in [11], no accuracy estimates are available.

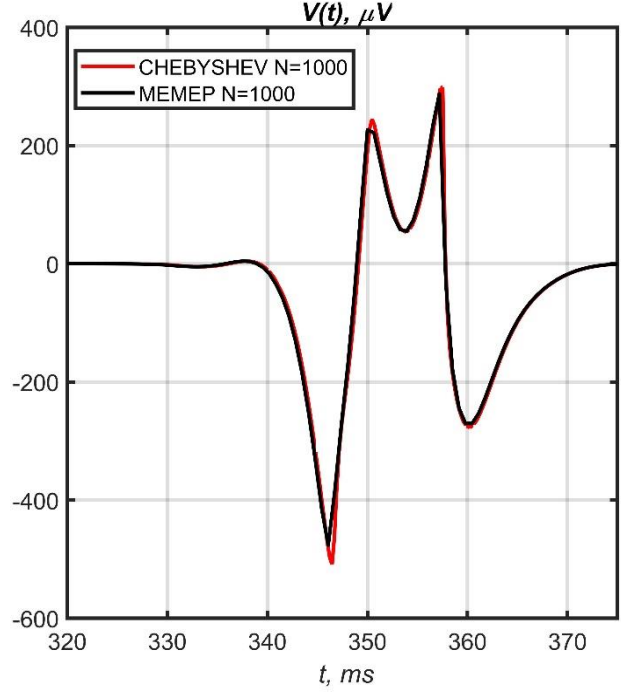

Figure 4. The benchmark problem. Solution by the Chebyshev polynomial-based method versus solution by the fastest finite element method in [11].

**Table 2.** Convergence of the two methods.

| Mesh points, N | Our method | | MEMEP | |
|---|---|---|---|---|
| | CPU time/cycle (s) | Relative error $\delta^N$ (%) | CPU time/cycle (s) | Relative error $\delta^N$ (%) |
| 100 | 1.2 | 2.3 | 11.5 | 0.85 |
| 200 | 4.8 | 0.67 | 45 | 0.29 |
| 400 | 33 | 0.18 | 155 | 0.12 |
| 1000 | 612 | - | 1030 | - |

## 5. Extension of the benchmark model

Our method is easily extended to the model with a field-dependent sheet critical current density, e.g.,

$$J_c = \frac{J_{c0}}{1+h_0^{-1}\sqrt{\kappa H_y^2 + H_z^2}}, \tag{16}$$

Where $J_{c0}, h_0, \kappa$ are fitting parameters. The parallel-to-film magnetic field component $H_y$, discontinuous on an infinitely thin film, is usually replaced by the corresponding component of the external field ($H_{\text{PM},y}$ in our case). To compute $H_z$ on each time step, the mesh values $J_k\sqrt{1-y_k^2}$ are interpolated by the Chebyshev expansion $\sum_{k=0}^{N-1}\alpha_k T_k(y)$ (Appendix A). This presents $J$ in the form (10); substituting (10) into equation (4) and using (12) one obtains

$$H_z(t, y_k) = H_{\text{PM},z}(\varphi(t), y_k) - \frac{1}{2}\sum_{i=1}^{N-1}\alpha_i U_{i-1}(y_k).$$

Finally, the mesh values of $J_c$ are calculated using (16) and the electric field values $E_k$ are determined by the dimensionless form of equation (2).

In our simulations (Figure 5) we chose $J_{c0}$ =236 A/cm equal to the constant $J_c$ in the benchmark problem, $\kappa = 0.5$, and three different values for $h_0$. For $h_0 = \infty$ we have $J_c \equiv J_{c0}$ and, the smaller $h_0$ is, the stronger is the dependence of the sheet critical current density on the magnetic field and the higher is the time-averaged voltage. Qualitatively, the voltage curves in Figure 5 are similar to the experimental ones, see [2, 7, 10].

As a model problem, an HTS dynamo connected to an electronically controlled current supply has been studied, both experimentally and numerically, in [10]. Numerical simulation in that work employed the $H$-formulation, and so a two-dimensional finite element mesh was needed in the plane orthogonal to the strip; the strip thickness has had to be artificially increased from 1 to 100 $\mu$m. A one-dimensional formulation, like the one used in our work, is expected to serve a basis for a more efficient numerical scheme.

Here we solve such a problem for several stator currents (see Figure 6). The field-dependent critical sheet current density (16) with $h_0 = 20 J_{c0}$ is assumed; initially, the stator current and the current densities in the strip are zero. The current $I$ was changed linearly with time during the first quarter of the first cycle, then remained a constant fraction of $I_c = 2aJ_{c0} = 283$ A. The mean dynamo generated power is $P = -I \cdot \langle V \rangle$, where $\langle V \rangle$ is the time averaged voltage (see Table 3). Our examples show that, depending on the applied current, this power can be positive (for $I = 0.1 I_c$ and $I = -0.2 I_c$) or negative (for $I = -0.1 I_c$). If the power is positive, the dynamo supplies energy to a load, otherwise the energy is consumed by the dynamo.

For the examples in this Section the computation for $N = 200$ took about six seconds per cycle and, as above, the solutions are shown for the second cycle.

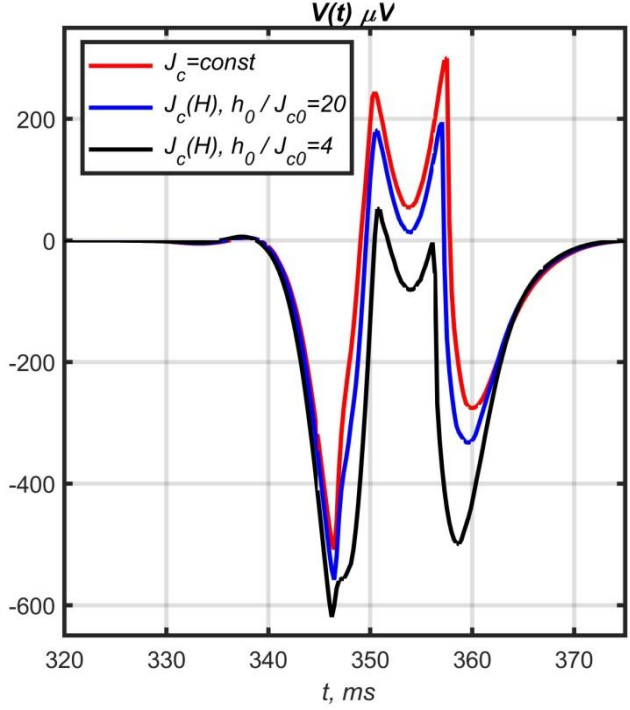 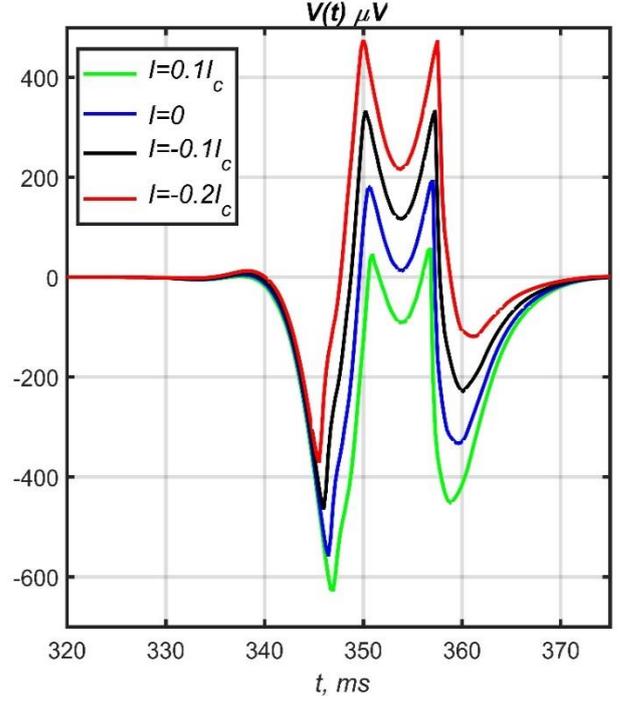

Figure 5. Simulation results for models with field-dependent critical sheet current density.

Figure 6. Simulation results for problems with transport current. Here the $J_c(H)$ dependence (16) is employed with $h_0 = 20 J_{c0}$.

**Table 3.** Mean voltage $\langle V \rangle$ during the second cycle for different stator currents.

| $I/I_c$ | 0.1 | 0 | -0.1 | -0.2 |
|---|---|---|---|---|
| $\langle V \rangle, \mu V$ | -25.7 | -17.8 | -5.1 | 10.8 |

## 6. Discussion

Devices for wireless injection of DC supercurrents into closed superconducting circuits open a way to design HTS magnets without thermally inefficient metal current leads. The HTS dynamo considered in our work is a device of this type; the 1D mathematical model [7-10] is able to describe, at least qualitatively, the main features of this device. In [11], this model was chosen as a benchmark problem and solved by a number of numerical methods; a comparison of the efficiency of these methods has been presented. Using expansions in Chebyshev polynomials for approximation in space and the method of lines for integration in time we derived a simple and accurate numerical method which is significantly faster than the methods considered in [11].

Extending the HTS dynamo open circuit benchmark problem, we solved also problems with field-dependent sheet critical current density and problems with transport current. Here we used the most often employed current-voltage dependence given by formulas (2), (16). However, this dependence can be easily replaced by, e.g., interpolation using the experimentally obtained data, as in [7, 8, 10]. Our approach is very efficient also for other superconductivity problems formulated as a singular integral equation of the Cauchy type or a system of such equations [12]. The method can be used also for mathematical modeling the traveling wave flux pumps [20] and multi-PM dynamos [5, 20].

## Appendix A. Operations with Chebyshev interpolating expansions

Our numerical scheme makes use of the following linear transformations: (*i*) from function values at the Chebyshev mesh (11) to coefficients of the interpolating Chebyshev polynomial expansion and vice versa; (*ii*) from the mesh values of the electric field $E$ to the approximate values of $\partial E / \partial y$ at the same mesh computed as derivatives of the interpolating polynomial; (*iii*) from the mesh values of $E$ to an approximation of its integral (6). Realization of these operations is described here for completeness.

(*i*) Let $\boldsymbol{g} = (g_1, ..., g_N)^T$ be the vector of function values at the mesh nodes. If $g(y) = \sum_{i=1}^{N} \gamma_i T_{i-1}(y)$ is the interpolating expansion in Chebyshev polynomials of the first kind, then $A\boldsymbol{\gamma} = \boldsymbol{g}$, where the $i$-th column of the $N \times N$ matrix $A$ is the vector $(T_{i-1}(y_1), ..., T_{i-1}(y_N))^T$. Efficient calculation of this matrix is based on the recurrent relation: $T_0 = 1$, $T_1 = y$, and $T_k = 2yT_{k-1} - T_{k-2}$ for $k \geq 2$. Multiplication by this matrix realizes transition from the expansion coefficients to function values; the inverse matrix is used for transition from function values to the expansion coefficients. Similarly, interpolation by the expansion in Chebyshev polynomials of the second kind, $g(y) = \sum_{i=1}^{N} \delta_i U_{i-1}(y)$, is related to the matrix $B$ with the $i$-th column $(U_{i-1}(y_1), ..., U_{i-1}(y_N))^T$; the matrix calculation is based on the relations: $U_0 = 1$, $U_1 = 2y$, and $U_k = 2yU_{k-1} - U_{k-2}$ for $k \geq 2$.

(*ii*) For a given vector $\boldsymbol{E}$ of $E_k = E(t, y_k)$ values, the derivatives $E'_k = \partial E / \partial y |_{t, y_k}$, see (14), are approximated by derivatives of the interpolating expansion $E = \sum_{i=1}^{N} \gamma_i T_{i-1}(y)$ with $\boldsymbol{\gamma} = A^{-1} \boldsymbol{E}$. Since $dT_k / dy = kU_{k-1}$, we have $E' = \sum_{i=2}^{N} (i-1) \gamma_i U_{i-2}(y)$. Let $C$ be the $N \times N$ matrix with elements

$$C_{ij} = \begin{cases} i & j = i+1 \\ 0 & j \neq i+1 \end{cases}.$$

Then $\boldsymbol{E'} = BC\boldsymbol{\gamma} = D\boldsymbol{E}$, where the $N \times N$ differentiation matrix $D = BCA^{-1}$.

(*iii*) Denote, for $i = 0, ..., N-1$,

$$\sigma_{i+1} = \int_{-1}^{1} T_i(y) dy = \begin{cases} \dfrac{2}{1-i^2} & i \text{ is even,} \\ 0 & i \text{ is odd.} \end{cases}$$

For $E = \sum_{i=1}^{N} \gamma_i T_{i-1}(y)$ we obtain $\int_{-1}^{1} E \, dy = \boldsymbol{\gamma}^T \cdot \boldsymbol{\sigma} = (A^{-1}\boldsymbol{E})^T \cdot \boldsymbol{\sigma} = \boldsymbol{E}^T \cdot (\boldsymbol{\sigma}^T A^{-1})^T = \boldsymbol{E}^T \cdot \boldsymbol{\eta}$, where $\boldsymbol{\sigma} = (\sigma_1, ..., \sigma_N)^T$ and $\boldsymbol{\eta} = (\boldsymbol{\sigma}^T A^{-1})^T$.

We note that matrices $A, A^{-1}, B, B^{-1}, D$ are used on each time step but, as well as the vector $\boldsymbol{\eta}$, should be calculated only once. For, e.g., $N = 200$ this takes about 0.01 second and ensures the efficiency of our simulations.

## Appendix B. Computing $\partial H_{PM,z} / \partial \varphi$: interpolation from a lookup table

We define a uniform mesh with $(N_\varphi + 1)$ nodes $\varphi_i$ in $[0, 2\pi]$ and find $H_{PM}$ analytically at the strip grid points $y_1, ..., y_N$ for every PM rotation angle $\varphi_i$ using (1). The lookup tables of $H_{PM,y}(\varphi_i, y_k)$ and $H_{\text{PM},z}(\varphi, y_k)$ are needed if the critical current density depends on magnetic field. The normal to strip component $H_{\text{PM},z}(\varphi, y_k)$ is, for each $y_k$, a periodic function of $\varphi$ sampled at the angles $\varphi_i$ and $\partial H_{\text{PM},z}(\varphi_i, y_k)/\partial \varphi$ values were obtained by numerical differentiation in the Fourier space using the fast Fourier transform. In our simulations we chose $N_\varphi = 2^{12}$; for $N = 200$ computing these tables takes about 0.1 second. Linear interpolation from the lookup tables provides a fast and accurate approximation to $H_{\text{PM},y}(\varphi, y_k)$, $H_{\text{PM},z}(\varphi, y_k)$, and $\partial H_{\text{PM},z}(\varphi, y_k)/\partial \varphi$ for any rotation angle $\varphi(t)$.

**ORCID iD**

Leonid Prigozhin  https://orcid.org/0000-0002-2448-4471
Vladimir Sokolovsky  https://orcid.org/0000-0003-4887-413X